\documentclass[journal, twoside, final]{IEEEtran}
\usepackage{cite}
\usepackage{url}
\usepackage{graphicx}
\usepackage{amssymb, amsmath, amsthm}
\usepackage{amsfonts}
\usepackage{multirow}
\usepackage{mathrsfs}
\usepackage{color}
\usepackage{flushend}
\usepackage{booktabs, multicol, multirow}
\usepackage{epsfig}
\allowdisplaybreaks

\newtheorem{proposition}{Proposition}

\begin{document}

\title{Modeling and Analysis of Cooperative Relaying in Spectrum-Sharing Cellular Systems}

\author{Minghua~Xia,~\IEEEmembership{Member,~IEEE}, and Sonia~A\"{\i}ssa,~\IEEEmembership{Senior Member,~IEEE}
\thanks{Copyright (c) 2015 IEEE. Personal use of this material is permitted. However, permission to use this material for any other purposes must be obtained from the IEEE by sending a request to pubs-permissions@ieee.org.}
\thanks{
Manuscript received June 25, 2015; revised October 26, 2015; accepted December 13, 2015. This work was supported by a Discovery Grant from the Natural Sciences and Engineering Research Council (NSERC) of Canada. The associate editor coordinating the review of this manuscript and approving it for publication was Prof. D. Zhao.}
\thanks{
M. Xia is with the School of Electronics and Information Engineering, Sun Yat-sen University (East Campus), Guangzhou, 510006, China (e-mail: xiamingh@mail.sysu.edu.cn). He was with the Institut National de la Recherche Scientifique (INRS), University of Quebec, Montreal, QC, H5A 1K6, Canada.}
\thanks{
S.~A\"{\i}ssa is with the Institut National de la Recherche Scientifique (INRS), University of Quebec, Montreal, QC, H5A 1K6, Canada (e-mail: aissa@emt.inrs.ca).}
\thanks{%
Color versions of one or more of the figures in this paper are available online at http://ieeexplore.ieee.org.}
\thanks{
Digital Object Identifier XXX}
}

\markboth{IEEE Transactions on Vehicular Technology, accepted for publication} {Xia \MakeLowercase{\textit{et al.}}: Modeling and Analysis of Cooperative Relaying in Spectrum-Sharing Cellular Systems}

\maketitle

\pubid{XXXX-XXXX~\copyright~2015 IEEE. Personal use is permitted, but republication/redistribution requires IEEE permission.}

\pubidadjcol

\begin{abstract}
\noindent In this paper, spectrum-sharing technology is integrated into cellular systems to improve spectrum efficiency. Macrocell users are primary users (PUs) while those within local cells, e.g., femtocell users, or desiring cost-effective services, e.g., roamers, are identified as secondary users (SUs). The SUs share the spectrum resources of the PUs in a underlay way, thus the transmit power of a secondary is strictly limited by the primary's tolerable interference power. Given such constraints, a cooperative relaying transmission between a SU and the macrocell base station (BS) is necessary. In order to guarantee the success of dual-hop relaying and avoid multi-hop relaying, a new cooperative paradigm is proposed, where an idle PU (instead of a secondary as assumed in general) in the vicinity of a target SU is chosen to serve as a relaying node, thanks to the fact that any PU can always transmit to the macrocell BS directly. Moreover, two-way relaying strategy is applied at the chosen relaying node so as to further improve the spectral efficiency. Our results demonstrate that the proposed system is particularly suitable for delay-tolerant wireless services with asymmetric downlink/uplink traffics, such as e-mail checking, web browsing, social networking and data streaming, which are the most popular applications for SUs in spectrum-sharing cellular networks.
\end{abstract}

\pubidadjcol

\section{Introduction}
\label{Section:Introduction}
\IEEEPARstart{I}{n} wireless environments, cognitive radio (CR) has a great potential to resolve the growing scarcity of the electromagnetic spectrum resources. Indeed, this technology allows secondary users (SUs) without explicitly assigned spectrum to coexist with primary users (PUs) licensed with particular spectrum. In general, there are three different schemes to implement CR, namely, underlay, overlay and interweaved \cite{Srinivasa07Mag05}. Among them, underlay CR, which is more commonly known as spectrum-sharing CR, does not involve complex spectrum-sensing mechanisms needed in interweaved CR or sophisticated encoding/decoding operation indispensable to overlay CR, and is particularly appealing in practical applications by enabling SUs to share spectrum resources of PUs as long as the harmful interference generated by SUs remains below pre-defined tolerable levels.

Given the advantages it brings in terms of spectrum utilization and efficiency, spectrum-sharing CR is highly appealing for integration in current and future wireless cellular systems such as IMT LTE-Advanced (4G). However, how to define SUs in primary cellular networks is still an open problem. Indeed, even if the distinction between users who should be licensed with dedicated spectrum resources and those who should access such resources in an opportunistic way is well defined, the maximum tolerable interference dictated by PUs confines the transmission between SUs to short-range communication. Only if the coverage of the secondary's transmission is extended can the application of spectrum-sharing techniques in cellular systems be made a reality and largely broadened. To this end, cooperative relaying techniques can be exploited. In particular, an idle user in the system can be leveraged to serve as a relaying node that assists the SU in transmitting to its far-end receiver, by avoiding interference levels that would otherwise result from a direct communication between the secondary and its destination and make the spectrum sharing with the PUs not feasible.

\pubidadjcol

In general, nodes acting as relays for a given secondary transmitter are always assumed to be other SUs available to assist the transmission in a dual-hop or a multi-hop manner. Under these settings, different relaying schemes have been studied in the open literature. For example, the performance of one-way decode-and-forward (DF) relaying in spectrum-sharing context was widely studied, see e.g., \cite{MusavianIET08, LeeTWC11Feb} and references therein. As well-known, the inherent decoding operation in DF relaying leads to higher implementation complexity and longer processing delay, compared to amplify-and-forward (AF) relaying. Recently, the performance of spectrum-sharing AF relaying was also studied, see e.g., \cite{AsghariICC10, XiaTCOM12June}. In particular, the effect of noise/interference amplification (or accumulation) inherent in dual-hop and multi-hop AF relaying was shown to yield significant degradation in the end-to-end performance of the secondary relaying link \cite{XiaWCL14}. Moreover, to avoid excessive interference at the PUs, hops along the secondary relaying link cannot work simultaneously, but rather in a consecutive way, which in turn causes degradation in spectrum efficiency. Due to these limitations, added to the above mentioned issue of the nature of SUs in primary cellular networks, the design and implementation of spectrum-sharing cooperative schemes in cellular systems is far from straightforward.

In order to improve the efficiency of spectrum utilization, in this paper we first propose a novel model to integrate spectrum-sharing technique into cellular systems by identifying potential SUs. Then, a new paradigm of cooperative relaying is proposed, where an idle PU (instead of a SU as generally assumed in the open literature) serves as relaying node to assist the data exchange between a SU and its target destination. Moreover, two-way relaying strategy is applied at the chosen relaying node so as to further improve the spectral efficiency.

By taking into account both the constraint on the tolerable interference power by PUs and the co-channel interference (CCI) originating from concurrent primary transmission, the outage probability at an active SU and at its target macrocell base station (BS) are analytically investigated. Our results disclose that the uplink performance of the considered secondary relaying link (from a SU to the macrocell BS) is dominated by the difference between the average tolerable interference power and the CCI, while the downlink performance (from the macrocell BS to the SU) depends mainly upon the average signal-to-interference ratio (SIR), regardless of the actual values of the tolerable interference power at PUs and the CCI. Due to its asymmetric downlink/uplink performance, the proposed scheme is particularly suitable for delay-tolerant wireless services with asymmetric downlink/uplink traffics, such as e-mail checking, web browsing, social networking and data streaming, which are most attractive to SUs.

The rest of this paper is organized as follows. Section \ref{Section:Modeling} describes the principle of the proposed primary/secondary spectrum-sharing model. Section \ref{Section:SystemModel} presents the signal model of the proposed relaying scheme and the optimal power allocation at the secondary. Section~\ref{Section:OutageAnalysis} analyzes the system performance in terms of the received SIRs at a SU and at its target BS.  Simulation results and discussions are presented in Section \ref{Section:SimulationResults}. Concluding remarks are provided in Section \ref{Section:Conclusion} and, finally, some detailed mathematical derivations are relegated to the appendix.
\section{Modeling of Cooperative Relaying in Spectrum-Sharing Cellular Systems}
\label{Section:Modeling}
\subsection{Who are Willing to be SUs in Cellular Systems?}
In order to improve the spectrum efficiency, CR technology is extensively believed to be applied in future cellular systems, yielding universal frequency reuse. However, how to integrate CR techniques into cellular systems is still an open issue. Actually, in current cellular networks, subscribers of a network operator get access to particular portions of licensed spectrum resources when needed and, thus, they are widely viewed as PUs from a CR point-of-view. Here, a question can be asked: who can, or is willing to, be secondary user in future cellular systems? Several scenarios can be envisioned. For instance, femtocell users underlaying in a macrocell \cite{AndrewsJSAC1403} are potential SUs because of their shorter transmission distance and lower transmit power relative to macrocell users and hence, low interference that they may inflict onto nearby macrocell users. Another example goes to roaming users. An obvious fact, as per popular accounting policy, is that the service fees for roamers are much more expensive than those of local users. As a result, many users keep their mobile terminals disconnected when they are out of the coverage area of their subscription network, which is not only undesirable but goes against the vision of ubiquitous and economically affordable wireless cellular access.

To allow the coexistence of SUs with PUs and guarantee no harmful impact on the quality-of-service (QoS) of macrocell PUs, several design criteria have to be developed. In particular, it is not hard to observe that the transmit power of SUs is generally lower than that of PUs due to {\it i)} their shorter transmission distance when they fall within a femtocell coverage and are serviced by its BS, or {\it ii)} the limitation on the interference power that can be tolerated by nearby macrocell PUs when the SUs are out of the femtocell BS coverage.\footnote{Strictly speaking, even when a SU is located inside a femtocell and serviced by its corresponding BS, the transmit power of the SU should also be limited by the tolerable interference power dictated by nearby PUs. However, in such a case, the SU's transmit power is usually very low and has little effect on nearby macrocell PUs who generally have relatively larger transmit power, by recalling the fact that the radius of femtocell coverage is only on the order of 10m whereas that of macrocell is about 500m.} Such low transmit power can be of no significant consequences for SUs located within the coverage of femtocells or in the vicinity of the macrocell BS and scheduled by the latter for service, given that they can communicate directly with the femtocell BSs or the macrocell BS as long as their QoS is satisfied.

\begin{figure}[t]
\centering
\includegraphics [width=3.5in, clip, keepaspectratio]{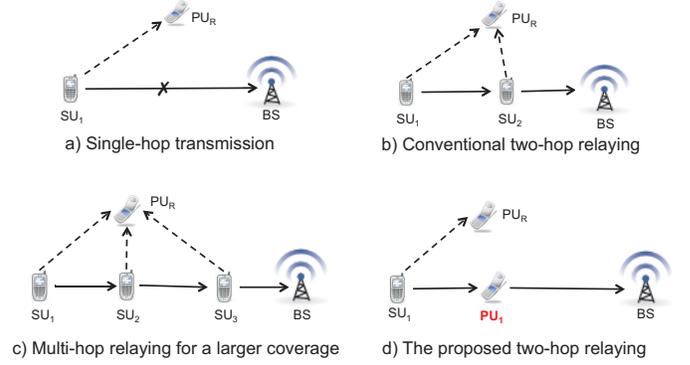}
\caption{A new paradigm of cooperative relaying where an idle PU acts as the relaying node between a source SU and the target macrocell BS.}
\label{Fig.RelayingModel}
\end{figure}

\subsection{Necessity of Relaying Between SUs and the Macrocell BS}
On the other hand, for SUs out of femtocell coverage and far from the macrocell BS, e.g., around the cell edge, they cannot communicate directly with the BS due to their strictly limited transmit power (cf. Fig.~\ref{Fig.RelayingModel}-a), which can result in significant service starvation. In such a case, cooperative relaying techniques can be exploited to enable them to communicate indirectly with the macrocell BS.\footnote{When a SU is out of the coverage of femtocells, it should in general transmit to the macrocell BS instead of a nearby femtocell BS. This is because most femtocells are of closed access and with limited capacity, e.g., serving only around one to four users \cite{ChandrasekharCM0809}. A closed access femtocell implies that it has a fixed set of domestic subscribers that, for privacy and security, are authorized to access the femtocell.} Conventionally, only a SU, instead of a PU, would assist another SU in transmitting to the macrocell BS. Since the transmit power of any SU, no matter the source or the relaying node along a secondary relaying link, is always strictly limited, two-hop (cf. Fig.~\ref{Fig.RelayingModel}-b) or even multi-hop relaying (cf. Fig.~\ref{Fig.RelayingModel}-c) is necessary for a successful data transfer from a SU to its target BS. However, the effect of noise/interference accumulation inherent in multi-hop AF relaying can significantly degrade the end-to-end performance of the secondary's relaying link \cite{XiaWCL14}. In particular, in order to avoid excessive interference at PUs, each hop along the multi-hop link cannot work simultaneously, but rather in a consecutive way. More specifically, if $K$ relaying nodes are involved, this will introduce $K+1$ transmission phases for a single data transfer between a source SU and the macrocell BS and, hence, will decrease the achievable data rate to $1/(K + 1)$, compared to that of the single-hop link (i.e., when $K = 0$).

Finally, a special case that may occur in practice is that some SUs falling within a femtocell coverage are refused to be serviced by the femtocell BS because of privacy and security. For these SUs, they can be treated in a similar way as below, except that their transmit power is limited by the minimum between the tolerable interference power imposed by PUs and that by the femtocell BS.
\subsection{How to Guarantee the Success of Dual-Hop Relaying Between SUs and the Macrocell BS?}
In order to address the aforementioned deficiencies of multi-hop relaying, we propose to rely on an {\it idle PU} to assist the data transfer between a SU and its macrocell BS, when the direct link is unreliable due to the secondary's limited transmit power, as illustrated in Fig.~\ref{Fig.RelayingModel}-d. The biggest advantage of the proposed scheme is that the SU can always reach its target BS within only two hops, thanks to the fact that any PU can always reach its target BS within a single hop, due to its relatively large transmit power and the capability of dynamically adjusting it. On the other hand, if the BS that a SU is originally assigned to is overloaded and cannot handle request anymore, the SU can leverage a nearby PU in an adjacent cell as a relaying node to communicate with the neighbouring BS. In this way, relaying techniques can be exploited not only to enhance the spectral efficiency for SUs but also to increase their chance of getting service, thus leading to higher overall network utilization efficiency with coexisting PUs and SUs. Furthermore, it is well-known that two-way relaying strategy yields higher spectral efficiency than the one-way counterpart. Therefore, in this paper we propose to exploit an idle PU and let it serve as a two-way relaying node between a SU and its target BS. The performance of the proposed two-way relaying link will be analytically investigated in the sequel of this paper.

A challenging question to the proposed two-way relaying model is why an idle PU would be willing to contribute to the data transfer of a SU? Actually, although PUs have already been compelled by telecommunications regulators like the Federal Communications Commission (FCC) to share their licensed spectrum resources with SUs, many system operators are reluctant to do so because of the lack of immediate compensation. To address this concern, it is critical to design some incentive mechanisms to encourage PUs to cooperate with SUs, ranging from technical to management perspectives \cite{Xia13Mag12, NTIA11,PCAST12}. For instance, a network operator can establish billing models that provide PUs who contribute to the cooperative scheme with service discount fees or credits. Billing and incentives models are beyond the scope of this paper, but it is evident that through smart billing strategies, any possible reduction in the average revenue per PU would not result in a loss for the operator but rather additional revenues from the increasing number of SUs, or new primary subscribers interested in cost-effective services.

Given the dynamics of the primary network in a spectrum-sharing environment, in terms of user distribution, the bursty nature of their traffic, and their willingness to cooperate according to pre-established billing and incentive models, it is reasonable to assume that the network operator would always be able to identify ideal PUs available for the aforementioned cooperation.

To improve the robustness of secondary transmission, a SU may firstly identify a candidate set consisting of several idle PUs. Then, the SU chooses an idle PU from the candidate set to serve as a relaying node, as per a certain criterion, for example, the idle PU with the shortest distance to the SU is firstly chosen. If this PU becomes active during secondary transmission, the SU stops transmitting signals via this PU and, then, chooses the PU with the second shortest distance to the SU from the candidate set to relay its data transmission. If there is no more idle PU available in the set, the SU has to suspend its transmission and wait until that a new non-empty candidate set is established. The way to identify an idle PU is similar to sensing a spectrum hole in cognitive radio context, but is beyond the scope of this paper.

On the other hand, if there are multiple SUs which concurrently want to relay signals via a same PU, the PU can choose a SU to serve according to their different priorities assigned by network operators as per, e.g., different amount of service fees that SUs have paid, or to a basic criterion like ``first come first serve'' if these SUs have the same priority.
\subsection{Possible Incentive Mechanisms for Idle PUs to Assist SUs}
In the state-of-the-art of research on CR systems, PUs are assumed to share their spectrum resources with SUs and tolerate some extra interference originating from SUs, so as to attain higher spectral efficiency compared to the conventional exclusive utilization of the spectrum resources. In this paper, we go one step further and assume that some idle PUs may serve as relays to assist the communication process between SUs and their target BS, in return of some revenue incentives, priority privileges or better QoS when needed.

Actually, from the management point of view, some spectrum authorities like the FCC and the National Telecommunications and Information Administration (NTIA) in the U.S. are considering new incentive mechanisms to promote more efficient use of the spectrum resources. For instance, the incentives subcommittee of spectrum management advisory committee of the NTIA suggested the FCC and the NTIA imposing license fees over all spectrum users and creating a innovation fund. The aim of the fund is to reimburse licensed spectrum users for their upfront research, planning and cooperation, among other costs \cite{NTIA11}.

In July 2012, the President's Council of Advisors on Science and Technology (PCAST) suggested the U.S. government opening up 1,000 MHz of Federal spectrum to commercial entities and creating an accounting and incentive system to promote more effective Federal spectrum use through a new dynamic spectrum-sharing model \cite{PCAST12}. Hence, in order to further improve spectral efficiency and QoS of both types of users, why not encourage idle PUs to assist SUs in exchanging their data by offering some incentives, for example, lower service fees, higher priority in times of emergency, or better QoS in challenging conditions? This is exactly the starting point of the present paper and its resulting contributions, in terms of modeling, analysis and findings.

\section{Signal Model and Power Allocation}
\label{Section:SystemModel}
In this section, the signal model and preliminary assumptions of the proposed system are firstly introduced. Then, the criterion of optimal power allocation at a source SU is established and, finally, the value of optimal transmit power at the SU is explicitly determined.

\begin{figure}[t]
\centering
\includegraphics [width=3.45in, clip, keepaspectratio]{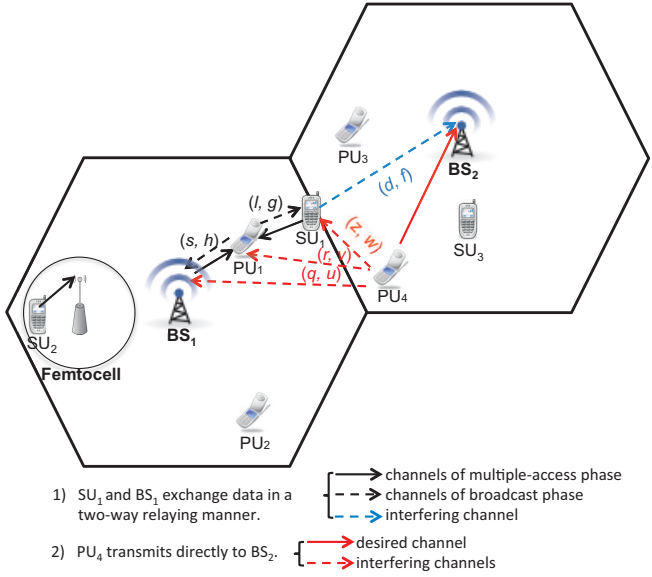}
\caption{The proposed two-way relaying scheme in spectrum-sharing cellular systems, where the primary user $\mathrm{PU}_1$ is exploited to assist the data exchange between the secondary user $\mathrm{SU}_1$ and the base station $\mathrm{BS}_1$.}
\label{Fig.SystemModel}
\end{figure}

\subsection{Signal Model}  \label{Section:SystemModel-ScenarioB}
As illustrated in Fig.~\ref{Fig.SystemModel}, a scheduled secondary user ($\mathrm{SU}_1$), is communicating with its target macrocell base station ($\mathrm{BS}_1$) through the assistance of a primary user ($\mathrm{PU}_1$) who acts as a two-way AF relay, while the primary user $\mathrm{PU}_4$ in an adjacent macrocell is transmitting to the neighbouring base station $\mathrm{BS}_2$. Accordingly, the received signals at the nodes $\mathrm{BS}_1$, $\mathrm{PU}_1$ and $\mathrm{SU}_1$ along the secondary relaying link are interfered by the CCIs coming from concurrent primary transmission originating from $\mathrm{PU}_4$, as indicated by the red arrowed dash lines in Fig.~\ref{Fig.SystemModel}.\footnote{It is remarkable that, when the reverse transmission, i.e., from $\mathrm{BS}_2$ to $\mathrm{PU}_4$,  is considered, the CCI originating from $\mathrm{BS}_2$ to the relaying link between $\mathrm{SU}_1$ and $\mathrm{BS}_1$ can be analyzed similarly in the way as below.}

Since two-way relaying strategy is applied, the communication process between $\mathrm{SU}_1$ and $\mathrm{BS}_1$ consists of two consecutive phases: multi-access~(MAC) phase and broadcast~(BC) phase. During the MAC phase, $\mathrm{SU}_1$ and $\mathrm{BS}_1$ simultaneously transmit signals to $\mathrm{PU}_1$. During the BC phase, $\mathrm{PU}_1$ amplifies its received sum signals with a power gain $\beta$ and broadcasts it to both $\mathrm{SU}_1$ and $\mathrm{BS}_1$ (how to determine the value of $\beta$ will be detailed later).

During the MAC phase, $\mathrm{BS}_1$ and $\mathrm{SU}_1$ transmit signals $x_1$ and $x_2$ with powers $P$ and $P_{\mathrm{su}_1}$ to the relay, respectively. It is assumed without loss of generality that $x_1$ and $x_2$ have the same amplitude. Accordingly, the received signal at the relaying node $\mathrm{PU}_1$ is given by
\begin{eqnarray}
y_{\mathrm{pu}_1}
& =  & \sqrt{P s^{-\epsilon}}\,h{x_1} + \sqrt{P_{\mathrm{su}_1} l^{-\epsilon}}\,g{x_2}  \nonumber \\
&     &{}+ \sqrt{P r^{-\epsilon}}\,v{x_3} + n_{\mathrm{pu}_1},
\label{Eq.Signal@Relay}
\end{eqnarray}
where $x_3$ refers to the interfering signal coming from concurrent primary transmitter $\mathrm{PU}_4$ and it has the same amplitude as $x_1$ and $x_2$, and where the channel parameter pair $(s, h)$ consists of the Euclidean distance, $s$, and the multi-path fading coefficient, $h$, between $\mathrm{BS}_1$ and $\mathrm{PU}_1$; the pairs $(l, g)$ and $(r, v)$ are defined similarly; $\epsilon \ge 2$ refers to the path-loss exponent, and $n_{\mathrm{pu}_1}$ denotes the additive white Gaussian noise (AWGN) at $\mathrm{PU}_1$ with zero mean and variance~$\sigma^2$. Also, $g$, $h$ and $v$ are supposed to be subject to independent and identically distributed (i.i.d.) Rayleigh block flat fading. That is, the values of $g$, $h$ and $v$ remain invariant during each data exchange between $\mathrm{BS}_1$ and $\mathrm{SU}_1$ but vary between two consecutive data exchanges.

During the BC phase, the relaying node $\mathrm{PU}_1$ amplifies its received signal, $y_{\mathrm{pu_1}}$, with a power gain $\beta$ ($\beta \in \Re_{+}$) and broadcasts it to $\mathrm{BS}_1$ and $\mathrm{SU}_1$. Accordingly, by taking into account the interfering signals coming from $\mathrm{PU}_4$, the received signals at $\mathrm{BS}_1$ and at $\mathrm{SU}_1$ are given by
\begin{equation}
y_{\mathrm{bs}_1}
= \sqrt{P s^{-\epsilon}}\,{h}{\beta}\,y_{\mathrm{pu}_1} + \sqrt{P q^{-\epsilon}}\,u{x_3} + n_{\mathrm{bs}_1}
\label{Eq.Signal@BS}
\end{equation}
and
\begin{equation}
y_{\mathrm{su}_1}
= \sqrt{P l^{-\epsilon}}\,{g}{\beta}\,y_{\mathrm{pu}_1} + \sqrt{P z^{-\epsilon}}\,w{x_3}+n_{\mathrm{su}_1},
\label{Eq.Signal@SU}
\end{equation}
respectively, where the parameter pairs $(q, u)$ and $(z, w)$ pertaining to the interfering channels from $\mathrm{PU}_4$ to $\mathrm{BS}_1$ and from $\mathrm{PU}_4$ to $\mathrm{SU}_1$, respectively, are defined in a similar way to the aforementioned channel parameters $(r, v)$ (cf. Fig.~\ref{Fig.SystemModel}). Then, substituting the relay gain $\beta$, defined as \cite{Nasab13TCOM08}
\begin{equation}
\label{Eq.PowerAmplifier}
\beta =  \left(Ps^{-\epsilon}|h|^2 + P_{\mathrm{su}_1}l^{-\epsilon}|g|^2 + Pr^{-\epsilon}|v|^2\right)^{-1/2},
\end{equation}
into \eqref{Eq.Signal@BS}-\eqref{Eq.Signal@SU}, subtracting the back-propagating self-interference \cite{RankovJSAC0702} and performing some algebraic manipulations, the received SIR at $\mathrm{BS}_1$ can be readily expressed as
\begin{equation}
\label{Eq.SIR@BS}
\gamma_{\mathrm{bs}_1} = \frac{\gamma_1\gamma_2}{\gamma_1+\gamma_2},
\end{equation}
where
\begin{equation}
\label{Eq.gamma1-2}
\gamma_1 = \eta_1 \frac{|h|^2}{|u|^2} ,\quad
\gamma_2 = \frac{P_{\mathrm{su}_1} l^{-\epsilon}|g|^2}{P\left(q^{-\epsilon}|u|^2 + r^{-\epsilon}|v|^2\right)},
\end{equation}
with $\eta_1 \triangleq (s/q)^{-\epsilon}$.
Similarly, the received SIR at $\mathrm{SU}_1$ is shown to be given by
\begin{equation} \label{Eq.SNR@SU}
\gamma_{\mathrm{su}_1} = \frac{\gamma_3\gamma_4}{\gamma_3+\gamma_5},
\end{equation}
where
\begin{equation}
\gamma_3  = \eta_2 \frac{|g|^2}{|w|^2},
\gamma_4  =  \eta_3 \frac{|h|^2}{|v|^2},
\gamma_5  =  \frac{P s^{-\epsilon} |h|^2+P_{\mathrm{su}_1} l^{-\epsilon} |g|^2}{P r^{-\epsilon} |v|^2}, \label{Eq.gamma3-5}
\end{equation}
with $\eta_2 \triangleq (l/z)^{-\epsilon}$ and $\eta_3 \triangleq (s/r)^{-\epsilon}$.

It is noteworthy that, because of their dedicated spectrum resources, the transmit powers at $\mathrm{BS}_1$ and at $\mathrm{PU}_1$ are fixed and identical to each other (i.e., $P$ as used before) by using a certain power control strategy prior to data transmission \cite{KnoppICC95}. On the other hand, for the secondary user $\mathrm{SU}_1$, in order not to inflict harmful interference on nearby PUs, its transmit power (i.e., $P_{\mathrm{su}_1}$) must be dynamically allocated, which will be elaborated below.

{\bf Remark 1 (The effect of noise variance on the received SINRs at both ends of the two-way relaying link)}:
If the noise variance is accounted for when computing the relay gain, \eqref{Eq.PowerAmplifier} can be rewritten as
\begin{equation}
\label{Eq.PowerAmplifier-b}
\beta^{\prime} =  \left(Ps^{-\epsilon}|h|^2 + P_{\mathrm{su}_1}l^{-\epsilon}|g|^2 + Pr^{-\epsilon}|v|^2 + \sigma^2\right)^{-1/2}.
\end{equation}
After some lengthy but straightforward mathematical manipulations, the received SINR at $\mathrm{BS}_1$ can be readily shown to be expressed as
\begin{equation}
\label{Eq.SIR@BS-b}
\gamma_{\mathrm{bs}_1} = \frac{\gamma_1\gamma_2}{\gamma_1+\gamma_2 + 1}.
\end{equation}
Clearly, when the product of $\gamma_1$ and $\gamma_2$ is large enough, \eqref{Eq.SIR@BS} is a tight upper bound of \eqref{Eq.SIR@BS-b}. Similarly, \eqref{Eq.SNR@SU} is a tight upper bound of the received SINR at $\mathrm{SU}_1$ if the relay gain is computed according to \eqref{Eq.PowerAmplifier-b}. Compared to \eqref{Eq.SIR@BS-b}, however, \eqref{Eq.SIR@BS} is much easier to be further processed due to its ease of mathematical tractability. As a result, it is \eqref{Eq.PowerAmplifier} rather than \eqref{Eq.PowerAmplifier-b} that is exploited to compute the relay gain throughout the paper.

\subsection{Criterion for Power Allocation at the SU}
\label{Subsection:Criteria4PowerAllocation-S2}
In order to maximize the achievable data rate at $\mathrm{BS}_1$ without inflicting harmful interference on PUs, the transmit power at $\mathrm{SU}_1$ should be optimized with respect to: {\it i)} the instantaneous interfering channel variations so as to satisfy the constraint on the tolerable interference power at its nearest primary receiver in the sense of minimum Euclidean distance, i.e., $\mathrm{BS}_2$ in Fig.~\ref{Fig.SystemModel} (cf.  the blue arrowed dash line from $\mathrm{SU}_1$ to $\mathrm{BS}_2$) and, {\it ii)} the CCI coming from the concurrent primary transmitter, i.e., $\mathrm{PU}_4$ in Fig.~\ref{Fig.SystemModel} (cf. the red arrowed dash line from $\mathrm{PU}_4$ to $\mathrm{SU}_1$). To this end, on one hand, it is clear that $\gamma_1$ involved in the received SIR $\gamma_{\mathrm{bs}_1}$ given by \eqref{Eq.SIR@BS} is independent of the transmit power at $\mathrm{SU}_1$. On the other hand, since the first-order derivative of $\gamma_{\mathrm{bs}_1}$ with respect to $\gamma_2$ is strictly positive, $\gamma_{\mathrm{bs}_1}$ in \eqref{Eq.SIR@BS} is a monotonically increasing function of $\gamma_2$. Hence, by virtue of the expression of $\gamma_2$ in \eqref{Eq.gamma1-2}, in order to maximize the achievable data rate at $\mathrm{BS}_1$, the optimal transmit power $P_{\mathrm{su}_1}$ at $\mathrm{SU}_1$ is determined as the solution to the optimization problem:\footnote{In general, for two-way relaying, the objective of dynamic power allocation at an end user is to maximize the sum rate achievable at both users. In this work, however, as it will be shown in Section~\ref{Subsection:Downlink} below, the received SIR at $\mathrm{SU}_1$ is not sensitive to its own transmit power but approaches an upper bound in the medium and high SIR regions. Therefore, we consider only the maximization of the achievable data rate at $\mathrm{BS}_1$.}
\begin{equation}  \label{Eq.CapacityA}
C = \max\limits_{P_{\mathrm{su}_1} \ge 0}\mathcal{E}_{g, u, v}
\left\{\log_2\left(1+\frac{P_{\mathrm{su}_1} l^{-\epsilon}|g|^2}{P\left(q^{-\epsilon}|u|^2 + r^{-\epsilon}|v|^2\right)}\right)\right\}
\end{equation}
\begin{equation}  \label{Eq.CapacityB}
\mathrm{s.t.} \quad \mathcal{E}_{f}\left\{P_{\mathrm{su}_1}d^{-\epsilon} |f|^2\right\} \le 10^{W/10},
\end{equation}
where the operator $\mathcal{E}_{x}\{y(x)\}$ means mathematical expectation of function $y(x)$ with respect to variable $x$; $W$ in the unit of dB with respect to the noise power denotes the average tolerable interference power at the nearest primary receiver.

Applying the Lagrangian optimization technique to (\ref{Eq.CapacityA})-(\ref{Eq.CapacityB}) in a similar way to \cite[Section 5.3.3]{Tse05}, it is easy to show that the optimal transmit power at $\mathrm{SU}_1$ is given by
\begin{equation} \label{Eq.CapacityC}
P_{\mathrm{su}_1} = \left[\frac{\lambda}{d^{-\epsilon}|f|^2} - \frac{P\left(q^{-\epsilon} |u|^2 + r^{-\epsilon}|v|^2\right)}{l^{-\epsilon} |g|^2}\right]^{\dag},
\end{equation}
where the ceiling operator $[x]^{\dag} \triangleq \max(0, x)$, and the power allocation parameter $\lambda$ in \eqref{Eq.CapacityC} is determined by the average interference power constraint satisfying the equality in (\ref{Eq.CapacityB}), such that
\begin{equation} \label{Eq.CapacityD}
\mathcal{E}_{f, g, u, v}\left\{\left[\lambda - P\left(q^{-\epsilon}|u|^2 + r^{-\epsilon}|v|^2\right)\frac{d^{-\epsilon}|f|^2}{l^{-\epsilon}|g|^2}\right]^{\dag}\right\} = 10^{\frac{W}{10}}.
\end{equation}

Like the well-known water-filling power allocation algorithm \cite{GoldsmithTIT97Nov}, the power allocation parameter $\lambda$ associated with \eqref{Eq.CapacityD} corresponds to the so-called water-level and it will be explicitly determined in the next subsection. The ceiling operator $[x]^{\dag}$ in (\ref{Eq.CapacityD}) implies that the transmit power is zero if the gain of the desired channel is smaller than or equal to a lower bound, i.e., $l^{-\epsilon}|g|^2 \le \frac{P}{\lambda}\left(q^{-\epsilon}|u|^2 + r^{-\epsilon}|v|^2\right)d^{-\epsilon}|f|^2$. In such a case, no data will be transmitted at the SU. This makes the two-way relaying in spectrum-sharing context completely different from and, in particular, more energy-efficient than conventional two-way relaying, where the transmission between two nodes is irrespective of the channel fluctuations in between. Moreover, it is observed that the aforementioned lower bound is determined by the product of the gain of the interfering channel, i.e., $d^{-\epsilon}|f|^2$ and the strength of the CCI, i.e., $q^{-\epsilon}|u|^2 + r^{-\epsilon}|v|^2$, which means that the impacts of the constraint on the tolerable interference power imposed by PUs and of the CCI coming from concurrent primary transmission, on the optimal power allocation at the SU are exchangeable.

Notice that, for two-way relaying without spectrum-sharing, the optimal power allocation in terms of maximizing the sum rate is equivalent to that of minimizing the outage probability \cite{XuTVT14}. However, in the context of spectrum-sharing, i.e., when the transmit power of SUs is strictly limited by the tolerable interference power dictated by PUs, as shown in \eqref{Eq.CapacityB}, more research efforts are needed to check the effectiveness of the aforementioned equivalence, which is beyond the scope of the paper.

{\bf Remark 2 (On the availability of channel state information (CSI) needed when performing power allocation at $\boldsymbol{\mathrm{SU}_1}$)}:
It is noteworthy that the channel parameters $(d, f)$, corresponding to the interfering channel from $\mathrm{SU}_1$ to $\mathrm{BS}_2$ as shown in Fig. \ref{Fig.SystemModel}, can be obtained at $\mathrm{SU}_1$ through periodic sounding of pilot signals transmitted by $\mathrm{BS}_2$. The channel parameters $(q, u)$ from $\mathrm{PU}_4$ to $\mathrm{BS}_1$ and $(r, v)$ from $\mathrm{PU}_4$ to $\mathrm{PU}_1$ can be obtained at $\mathrm{SU}_1$ through the feedback from $\mathrm{BS}_1$ and $\mathrm{PU}_1$, respectively. Although the acquisition of these CSI requires additional cost at $\mathrm{SU}_1$, it enables $\mathrm{SU}_1$ to strictly comply with the interference power constraint dictated by PUs and to maximize its achievable data rate.

{\bf Remark 3 (Constraints on the transmit power of SUs)}:
In CR systems, the tolerable interference power at PUs can be generally defined by means of average interference power or peak interference power or both \cite{MusavianTWC09Jan}. The average interference power constraint for SUs applies to non-real time applications and has low feedback overhead. The peak interference power constraint for SUs is suitable for real-time applications and has high feedback overhead. Also, there is a maximum output-power constraint for SUs in practice, i.e., a physically allowable maximum transmit power. Its effect on system performance is essentially equivalent to peak interference power constraint mentioned earlier. Nevertheless, it is demonstrated that ``imposing a constraint on the peak interference power does not yield a significant impact on the ergodic capacity as long as the average interference power is constrained'' \cite{MusavianTWC09Jan}. Hence, only the average interference power constraint is considered in the optimization problem formulated in \eqref{Eq.CapacityA}-\eqref{Eq.CapacityB}. For more detailed comparison between average and peak interference power constraints, the interested reader is referred to \cite{Zhang09TWC04}.

\subsection{Optimal Transmit Power at the SU}
\label{Subsection:PowerAllocation}
Now, we explicitly derive the value of the power-allocation parameter $\lambda$ in \eqref{Eq.CapacityD} and, then, it is applied to \eqref{Eq.CapacityC} to determine the optimal transmit power at $\mathrm{SU}_1$. To this end, we define a new random variable $T \triangleq \left(q^{-\epsilon}|u|^2 + r^{-\epsilon}|v|^2\right)\frac{|f|^2}{|g|^2} = V_1V_3$, where $V_1 \triangleq |f|^2/|g|^2$ and $V_3 \triangleq q^{-\epsilon}|u|^2 + r^{-\epsilon}|v|^2$. By recalling that the multi-path fading components of all channels in the considered system are supposed to be subject to Rayleigh fading, the PDFs of $|f|^2$ and $|g|^2$ are of the same exponential distribution:
\begin{equation}  \label{Eq.PDFRayleigh}
f_X(x) = \frac{1}{\bar{\gamma}}\exp\left(-\frac{x}{\bar{\gamma}}\right), \quad  X \in \left\{|f|^2, \, |g|^2\right\}
\end{equation}
where $\bar{\gamma}$ is the average SNR of the signals transmitted over the channel, provided that the average power gain of the considered channel is normalized,

In light of \eqref{Eq.PDFRayleigh} and conditioning on $|g|^2$, the PDF of $V_1$ can be easily given by
\begin{equation}
\label{Eq.PDFV}
f_{V_1}(x) = (x+1)^{-2}.
\end{equation}
On the other hand, it is clear that $V_3$ is the sum of two exponentially distributed variables with mean $q^{-\epsilon}$ and $r^{-\epsilon}$, respectively. Thus, as per \cite[Eqs.(18.28)--(18.29)]{Balakrishnan03}, the PDF of $V_3$ can be readily given by
\begin{equation} \label{Eq.PDFV3}
f_{V_3}(x) =
\begin{cases}
c_1\left(e^{-q^{\epsilon}x} - e^{-r^{\epsilon}x}\right), & \text{if } q \ne r; \\
\frac{x}{q^{2\epsilon}} \, e^{-\frac{x}{q^{\epsilon}}}, & \text{if } q = r,
\end{cases}
\end{equation}
where $c_1 \triangleq 1/\left(q^{-\epsilon}-r^{-\epsilon}\right)$. In practical cellular communication systems, the distance parameters $q$ and $r$ are generally unequal to each other (cf. Fig.~\ref{Fig.SystemModel}) and, thus, in the following we concentrate on the upper case of \eqref{Eq.PDFV3}. If the lower case with $q = r$ in \eqref{Eq.PDFV3} is to be considered, it can be analyzed in a similar way.

Specifically, by virtue of \eqref{Eq.PDFV} and \eqref{Eq.PDFV3}, the CDF of $T$ can be derived as follows.
\begin{align}
F_{T}(x)
&  =   \int_0^\infty{F_{V_1}\left(\frac{x}{y}\right) f_{V_3}(y)}\,\mathrm{d}y  \nonumber \\
&  =   c_1\int_0^\infty{\frac{x}{x+y}\left(e^{-q^{\epsilon}y} - e^{-r^{\epsilon}y}\right)}\,\mathrm{d}y   \label{Eq.CDFT-1}\\
&  =   c_1 x\left[\Psi\left(1,\,1,\, q^{\epsilon}x\right) - \Psi\left(1,\,1,\, r^{\epsilon}x\right) \right]
\label{Eq.CDFT-2} \\
&  =   \frac{c_1}{q^{\epsilon}} G_{1,\,2}^{2,\,1}\left[q^{\epsilon}x \left\vert \begin{gathered} 1 \\ 1, 1 \end{gathered} \right.\right]
 - \frac{c_1}{r^{\epsilon}} G_{1,\,2}^{2,\,1}\left[r^{\epsilon}x \left\vert \begin{gathered} 1 \\ 1, 1 \end{gathered} \right.\right],
\label{Eq.CDFT}
\end{align}
where $G[\,x\, \vert\,.\,]$ denotes the Meijer's $G$-function \cite[Eq.(16.17.1)]{NIST10}, and \cite[vol.1, Eq.(2.3.6.9)]{Prudnikov86} was exploited to reach \eqref{Eq.CDFT-2} and \cite[vol.3, Eq.(8.4.46.1)]{Prudnikov86} was used to attain \eqref{Eq.CDFT}. Moreover, in light of \cite[vol.3, Eq.(8.2.2.32)]{Prudnikov86} and by taking the derivative of \eqref{Eq.CDFT} with respect to $x$, we obtain the PDF of $T$, given by
\begin{equation} \label{Eq.PDFT}
f_{T}(x)
=   \frac{c_1}{x \, q^{\epsilon}} G_{2,\,3}^{2,\,2}\left[q^{\epsilon}x \left\vert \begin{gathered} 0, 1 \\ 1, 1, 1 \end{gathered} \right.\right]
 - \frac{c_1}{x \, r^{\epsilon}} G_{2,\,3}^{2,\,2}\left[r^{\epsilon}x \left\vert \begin{gathered} 0, 1 \\ 1, 1, 1 \end{gathered} \right.\right].
\end{equation}

Subsequently, in light of  \eqref{Eq.CapacityD} and \eqref{Eq.PDFT}, the power-allocation parameter $\lambda$ can be determined by
\begin{align}
10^{\frac{W}{10}}
&  =    \int_0^{\frac{\lambda}{\eta_4\bar{\gamma}P}}\left(\lambda-{\eta_4}P x\right)f_{T}(x)\,\mathrm{d}x        \nonumber \\
&  =    \lambda F_{T}\left(\frac{\lambda}{\eta_4\bar{\gamma}P}\right)
- {\eta_4}P \int_0^{\frac{\lambda}{\eta_4\bar{\gamma}P}}xf_{T}(x)\,\mathrm{d}x,	 \label{Eq.PowerAllocationCCI-1}
\end{align}
where $\eta_4 \triangleq (d/l)^{-\epsilon}$. Then, applying the integration-by-parts method to the second term of \eqref{Eq.PowerAllocationCCI-1} and performing some mathematical manipulations with the help of \eqref{Eq.CDFT}, we eventually obtain \eqref{Eq.PowerAllocationCCI-2} at the top of the next page.

With the value of $\lambda$ numerically established as per \eqref{Eq.PowerAllocationCCI-2}, the optimal transmit power $P_{\mathrm{su}_1}$ at $\mathrm{SU}_1$ can be readily determined by substituting it into \eqref{Eq.CapacityC}. Then, with the resultant $P_{\mathrm{su}_1}$, we derive the distribution functions of the received SIRs at $\mathrm{BS}_1$ and at $\mathrm{SU}_1$ in the next section.

\newcounter{mytempeqncnPOWER}
\begin{figure*}[!t]
\normalsize
\setcounter{mytempeqncnPOWER}{\value{equation}}
\setcounter{equation}{22}
\begin{eqnarray}  \label{Eq.PowerAllocationCCI-2}
10^{\frac{W}{10}}
&  =  & c_1 \lambda\left(1-\frac{1}{\bar{\gamma}}\right)\left\{q^{-\epsilon}G_{1,\,2}^{2,\,1}\left[ \frac{\lambda}{\eta_4\bar{\gamma}Pq^{-\epsilon}} \left\vert \begin{gathered} 1 \\ 1, 1 \end{gathered} \right.\right]
 - r^{-\epsilon}G_{1,\,2}^{2,\,1}\left[ \frac{\lambda}{\eta_4\bar{\gamma}Pr^{-\epsilon}}\left\vert \begin{gathered} 1 \\ 1, 1 \end{gathered} \right.\right]\right\}   \nonumber \\
&      & {}+c_1{\eta_4}P\left\{q^{-2\epsilon}G_{2,\,3}^{2,\,2}\left[\frac{\lambda}{\eta_4\bar{\gamma}Pq^{-\epsilon}} \left\vert \begin{gathered} 1, 2 \\ 2, 2, 0 \end{gathered} \right.\right]
 - r^{-2\epsilon}G_{2,\,3}^{2,\,2}\left[\frac{\lambda}{\eta_4\bar{\gamma}Pr^{-\epsilon}} \left\vert \begin{gathered} 1, 2 \\ 2, 2, 0 \end{gathered} \right.\right]\right\}.
\end{eqnarray}
\setcounter{equation}{\value{mytempeqncnPOWER}}
\hrulefill
\vspace*{6pt}
\end{figure*}
\setcounter{equation}{23}

\section{Analysis of the End-to-End SIRs}
\label{PerformanceAnaylis}
In this section, we analyze the distribution functions of the end-to-end SIRs from $\mathrm{SU}_1$ to its target base station $\mathrm{BS}_1$ and from $\mathrm{BS}_1$ to the $\mathrm{SU}_1$, both through the relaying node $\mathrm{PU}_1$.

\label{Section:OutageAnalysis}
\subsection{The Received SIR at the BS}
\label{Subsection:Uplink-S2}
Here, we derive the distribution functions of the received SIR at the BS. By virtue of the received SNR $\gamma_{\mathrm{bs}_1}$ in (\ref{Eq.SIR@BS}), in order to determine its distribution functions, we need to firstly derive the distribution functions of its two components $\gamma_1$ and $\gamma_2$ shown in \eqref{Eq.gamma1-2}. By using a similar approach as in \eqref{Eq.PDFV}, the PDF and CDF of $\gamma_1$ can be easily expressed as
\begin{equation}  \label{Eq.PDFgamma1}
f_{\gamma_{_1}}(x) = \eta_1\bar{\gamma}\left(x+\eta_1\bar{\gamma}\right)^{-2}
\end{equation}
and
\begin{equation}  \label{Eq.CDFgamma1}
F_{\gamma_{_1}}(x) = 1-\eta_1\bar{\gamma}\left(x+\eta_1\bar{\gamma}\right)^{-1},
\end{equation}
respectively. On the other hand, substituting the value of the optimal transmit power given by \eqref{Eq.CapacityC} into the definition of $\gamma_2$ shown in \eqref{Eq.gamma1-2}, $\gamma_2$ can be reformulated as
\begin{equation}  \label{Eq.gamma2-R1}
\gamma_2 = \left[\frac{\lambda\, l^{-\epsilon}|g|^2}{P\left(q^{-\epsilon}|u|^2 + r^{-\epsilon}|v|^2\right) d^{-\epsilon}|f|^2}-1\right]^{\dag}
= \left[\frac{c_2}{T}-1\right]^{\dag},
\end{equation}
where the constant $c_2 \triangleq \lambda/({\eta_4}P)$. Subsequently, by performing some algebraic manipulations, the PDF and CDF of $\gamma_2$ can be readily expressed as
\begin{equation}  \label{Eq.PDFgamma2}
f_{\gamma_{_2}}(x) = \frac{c_2}{F_{T}(c_2) \, (x+1)^2} \, f_{T}\left(\frac{c_2}{x+1}\right)
\end{equation}
and
\begin{equation}  \label{Eq.CDFgamma2}
F_{\gamma_{_2}}(x) = 1-\frac{1}{F_{T}(c_2)}F_{T}\left(\frac{c_2}{x+1}\right),
\end{equation}
respectively, where $F_T(x)$ and $f_T(x)$ are explicitly defined in \eqref{Eq.CDFT} and \eqref{Eq.PDFT}, respectively.

Due to the high complexity of \eqref{Eq.PDFgamma2}-\eqref{Eq.CDFgamma2}, it is mathematically intractable to derive the exact distribution functions of the received SIR $\gamma_{\mathrm{bs}_1}$ given by \eqref{Eq.SIR@BS}. On the other hand, it is well-known that $\gamma_{\mathrm{bs}_1}$ can be bounded by \cite{AnghelTWC04May}
\begin{equation} \label{Eq.SIR@BS-bounds}
\frac{1}{2}\min\{\gamma_1,\,\gamma_2\} \le \gamma_{\mathrm{bs}_1} = \frac{\gamma_1\gamma_2}{\gamma_1+\gamma_2} \le \min\{\gamma_1,\,\gamma_2\}.
\end{equation}
Then, by recalling the result in the theory of order statistics, the CDF of $\gamma_{\mathrm{bs}_1}$ can be shown to be bounded by
\begin{eqnarray} \label{Eq.CDF@BS-bounds}
\lefteqn{F_{\gamma_1}(\gamma)+F_{\gamma_2}(\gamma) - F_{\gamma_1}(\gamma)F_{\gamma_2}(\gamma)}   \nonumber \\
& ~\le~    F_{\gamma_{\mathrm{bs}_1}}(\gamma)  ~\le~ &          \nonumber \\
&F_{\gamma_1}(2\gamma)+F_{\gamma_2}(2\gamma) - F_{\gamma_1}(2\gamma)F_{\gamma_2}(2\gamma).&
\end{eqnarray}

It is remarkable that only if $\gamma_1 = \gamma_2$ does the value of $\gamma_{\mathrm{bs}_1}$ equals the lower bound shown in \eqref{Eq.SIR@BS-bounds}. Actually, due to their different definitions as shown in \eqref{Eq.gamma1-2}, the values of $\gamma_1$ and $\gamma_2$ are quite different and, thus, the value of $\gamma_{\mathrm{bs}_1}$ approaches its upper bound shown in \eqref{Eq.SIR@BS-bounds}. As a result, by recalling the fact that the CDF of $\gamma_{\mathrm{bs}_1}$, i.e., $F_{\gamma_{\mathrm{bs}_1}}(\gamma)$, is a monotonically decreasing function with respect to $\gamma$, it is evident that $F_{\gamma_{\mathrm{bs}_1}}(\gamma)$ in \eqref{Eq.CDF@BS-bounds} should approach its lower bound. This observation will also be demonstrated by simulation results in the following Section~\ref{Subsection:Outage}.

\subsection{Upper Bound on the Received SIR at the SU}
\label{Subsection:Downlink}
Now, we turn to the received SIR at $\mathrm{SU}_1$, i.e., $\gamma_{\mathrm{su}_1}$ shown in \eqref{Eq.SNR@SU}. Since $P_{\mathrm{su}_1}$ given by \eqref{Eq.CapacityC} is involved in the definition of $\gamma_5$ given by \eqref{Eq.gamma3-5}, the distribution functions of $\gamma_5$ are mathematically intractable and, thus, the exact distribution functions of $\gamma_{\mathrm{su}_1}$ are not available. In order to proceed, an upper bound on $\gamma_{\mathrm{su}_1}$ is introduced. Specifically, in light of the definitions of $\gamma_4$ and $\gamma_5$ shown in \eqref{Eq.gamma3-5}, it is clear that $\gamma_4 \le \gamma_5$, where the equality holds only if the transmit power $P_{\mathrm{su}_1}=0$. Accordingly, the received SIR at $\mathrm{SU}_1$, given by \eqref{Eq.SNR@SU}, can be upper bounded by
\begin{equation} \label{Eq.SNR@SU-bound}
\gamma_{\mathrm{su}_1}
= \frac{\gamma_3\gamma_4}{\gamma_3+\gamma_5}
\le \underbrace{\frac{\gamma_3\gamma_4}{\gamma_3+\gamma_4}}_{\gamma^{\prime}_{\mathrm{su}_1}}.
\end{equation}
Furthermore, it is clear that $\gamma_3$ and $\gamma_4$ defined in \eqref{Eq.gamma3-5} are independent of each other and their PDF and CDF are of the same form as those of $\gamma_1$ (i.e., \eqref{Eq.PDFgamma1} and \eqref{Eq.CDFgamma1}), except that the parameter $\eta_1$ should be replaced by $\eta_2$ for the distribution functions of $\gamma_3$ (or $\eta_1$ replaced by $\eta_3$ for those of $\gamma_4$). As a result, the exact distribution functions of $\gamma^{\prime}_{\mathrm{su}_1}$ given by \eqref{Eq.SNR@SU-bound} can be derived and they are summarized in the following proposition.

\begin{proposition}  \label{Proposition1@SU}
The CDF and PDF of $\gamma^{\prime}_{\mathrm{su}_1}$ shown in \eqref{Eq.SNR@SU-bound} are given by \eqref{Eq.CDFSNR@SU-bound} and \eqref{Eq.PDFSNR@SU-bound}, respectively, where ${_2F_1}( \cdot\,, \cdot\,; \cdot\,; x)$ refers to the Gaussian hypergeometric function \cite[Eq.(15.2.1)]{NIST10}.
\end{proposition}

\begin{IEEEproof}
See the Appendix.
\end{IEEEproof}

\newcounter{mytempeqncntA-1}
\begin{figure*}[!t]
\normalsize
\setcounter{mytempeqncntA-1}{\value{equation}}
\setcounter{equation}{31}
\begin{eqnarray}
\label{Eq.CDFSNR@SU-bound}
F_{\gamma^{\prime}_{\mathrm{su}_1}}(x)
&  =  & 1 - \frac{\eta_2 \eta_3 \bar{\gamma}^2 x^2}{2(x+\eta_2\bar{\gamma})^2 (x+\eta_3\bar{\gamma})^2}
\,{_2F_1}\left(2, 2; 3; 1-\frac{x^2}{(x+\eta_2\bar{\gamma})(x+\eta_3\bar{\gamma})}\right)
\end{eqnarray}
\begin{eqnarray}
\label{Eq.PDFSNR@SU-bound}
f_{\gamma^{\prime}_{\mathrm{su}_1}}(x)
&  =  & \frac{\eta_2 \eta_3 \bar{\gamma}^2 x (x^2 - \eta_2 \eta_3 \bar{\gamma}^2)}{(x+\eta_2\bar{\gamma})^3 (x+\eta_3\bar{\gamma})^3}
\,{_2F_1}\left(2, 2; 3; 1-\frac{x^2}{(x+\eta_2\bar{\gamma})(x+\eta_3\bar{\gamma})}\right)   \nonumber \\
&      & {} + \frac{2\eta_2 \eta_3 \bar{\gamma}^2{x^3}\left[x(\eta_2\bar{\gamma}+\eta_3\bar{\gamma}) + 2\eta_2\eta_3\bar{\gamma}^2\right]}{3(x+\eta_2\bar{\gamma})^4 (x+\eta_3\bar{\gamma})^4}
\,{_2F_1}\left(3, 3; 4; 1-\frac{x^2}{(x+\eta_2\bar{\gamma})(x+\eta_3\bar{\gamma})}\right)
\end{eqnarray}
\setcounter{equation}{\value{mytempeqncntA-1}}
\hrulefill
\vspace*{6pt}
\end{figure*}
\setcounter{equation}{33}

\section{Simulation Results and Discussions}
\label{Section:SimulationResults}
In this section, the results obtained in the preceding sections are applied to analyze and gain insights into the system performance. As generally termed in cellular communication systems, the achievable performance on the BS side is referred to as {\it uplink} performance (corresponding to the link $\mathrm{SU}_1 \to \mathrm{PU}_1 \to \mathrm{BS}_1$ in Fig.~\ref{Fig.SystemModel}) and the performance on the SU side is known as {\it downlink} performance (corresponding to the link $\mathrm{BS}_1 \to \mathrm{PU}_1 \to \mathrm{SU}_1$ in Fig.~\ref{Fig.SystemModel}).

\subsection{Simulation Scenarios and Parameter Setting}
\begin{figure}[t]
\centering
\includegraphics [width=2.1in, clip, keepaspectratio]{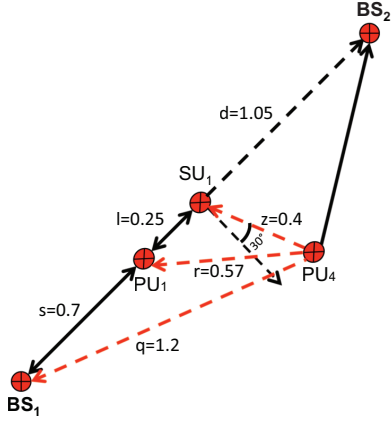}
\caption{Simulation setting with specific distance parameters, where macrocell radius is normalized.}
\label{Fig.SimulationSetting}
\end{figure}

Figure~\ref{Fig.SimulationSetting} illustrates a typical simulation scenario with specific distance parameters, where the geometry of all nodes and the distances among them are fixed, although they have different directions of data flow. In order to determine the geometry of all nodes without loss of generality, cell radius is normalized to unity. That is, the distance between base stations $\mathrm{BS}_1$ and $\mathrm{BS}_2$ is set to~$2$. Moreover, $\mathrm{PU}_1$ and $\mathrm{SU}_1$ are deployed along the segment between $\mathrm{BS}_1$ and $\mathrm{BS}_2$. On the other hand, the interfering channel from $\mathrm{PU}_4$ to $\mathrm{SU}_1$ forms an angle of $30^{\circ}$ with the line perpendicular to the segment between $\mathrm{BS}_1$ and $\mathrm{BS}_2$. Also, the normalized distance between secondary transmitter $\mathrm{SU}_1$ and its relaying node $\mathrm{PU}_1$ is set to $0.25$ and the distance between $\mathrm{SU}_1$ and the concurrent active primary transmitter $\mathrm{PU}_4$ is $0.4$. With these definitions in mind, other parameters can be determined accordingly and they are explicitly shown in Fig.~\ref{Fig.SimulationSetting}.

In the ensuing Monte-Carlo simulation experiments, the variance of AWGN noise ($\sigma^2$) at any node is set to unity. The CCI originating from $\mathrm{PU}_4$ is normalized with respect to the noise variance, which is essentially equal to the interference-to-noise ratio (INR) in the unit of dB. The interfering channels from $\mathrm{PU}_4$ to all the nodes along the secondary relaying link are subject to i.i.d. Rayleigh block flat fading with unit variance. On the other hand, the energy of each transmitted symbol at either $\mathrm{SU}_1$ or $\mathrm{BS}_1$ is scaled by the value of the average SIR. Moreover, the outage threshold at either the BS or $\mathrm{SU}_1$  is set to 3, i.e., $4.7712$dB with respect to normalized noise variance, which means that the minimum data-rate requirement without outage occurrence at either the BS or $\mathrm{SU}_1$ is set to $\frac{1}{2}\log_{2}(1+3)=1$ bit/s/Hz. This is probably a minimal requirement for a successful voice call \cite{Andrews11TCOM11}.

\subsection{Outage Probability Analysis}
\label{Subsection:Outage}
In this subsection, we apply the obtained results in Section~\ref{PerformanceAnaylis} to illustrate the outage probability performance of the proposed relaying scheme. In principle, outage probability is defined as the probability that the instantaneous received SIR falls below a pre-defined threshold value~$\gamma_\mathrm{th}$. In practice, outage probability can be readily evaluated by using the CDF of the received SIR. Specifically, in light of \eqref{Eq.CDF@BS-bounds}, the outage probability at $\mathrm{BS}_1$, $\mathrm{P}_{\mathrm{bs}_1}(\gamma{_\mathrm{th}})$, can be bounded by
\begin{eqnarray}  \label{Eq.Outage-BS}
\lefteqn{F_{\gamma_1}(\gamma{_\mathrm{th}})+F_{\gamma_2}(\gamma{_\mathrm{th}})
- F_{\gamma_1}(\gamma{_\mathrm{th}})F_{\gamma_2}(\gamma{_\mathrm{th}})}   \nonumber \\
& ~\le~    \mathrm{P}_{\mathrm{bs}_1}(\gamma{_\mathrm{th}})  ~\le~ &          \nonumber \\
&F_{\gamma_1}(2\gamma{_\mathrm{th}})+F_{\gamma_2}(2\gamma{_\mathrm{th}})
- F_{\gamma_1}(2\gamma{_\mathrm{th}})F_{\gamma_2}(2\gamma{_\mathrm{th}}).&
\end{eqnarray}

\begin{figure}[t]
\centering
\includegraphics [width=3.4in, clip, keepaspectratio]{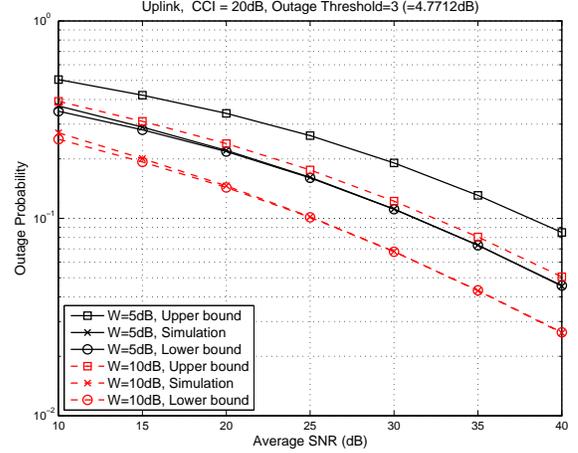}
\caption{Effectiveness of the lower bound shown in \eqref{Eq.Outage-BS} on outage probability of the received SIR at the base station.}  \label{Fig.OutageUplinkBounds}
\end{figure}

In order to demonstrate the effectiveness of the obtained bounds shown in \eqref{Eq.Outage-BS}, Fig.~\ref{Fig.OutageUplinkBounds} depicts this outage probability versus the average SIR in dB, where the CCI is fixed to $20$ dB whereas the average tolerable interference power $W$ varies from $5$dB to $10$dB. It is observed from Fig.~\ref{Fig.OutageUplinkBounds} that the lower bound computed by \eqref{Eq.CDF@BS-bounds} is very tight with the simulation results at the medium and high SNR, since in general the values of $\gamma_1$ and $\gamma_2$ defined in \eqref{Eq.gamma1-2} are quite different and, in turn, the received SIR approaches it upper bound given by \eqref{Eq.SIR@BS-bounds}. On the contrary, the upper bound is always very loose since, as per \eqref{Eq.SIR@BS-bounds}, only if the values of $\gamma_1$ and $\gamma_2$ are almost identical does the lower bound on the received SIR become tight, which is not the case in practice due to their different definitions.

On the other hand, Fig.~\ref{Fig.OutageUplinkBounds} shows that the outage probability at $\mathrm{BS}_1$ decreases with $W$, i.e., the average tolerable interference power at PUs. This is because, larger $W$ allows larger transmit power at $\mathrm{SU}_1$ and, subsequently, larger $\gamma_2$ as per \eqref{Eq.gamma1-2} and, finally, larger $\gamma_{\mathrm{bs}_1}$, since $\gamma_{\mathrm{bs}_1}$ is a monotonically function with respect to $\gamma_2$ according to \eqref{Eq.SIR@BS}.

\begin{figure}[t]
\centering
\includegraphics [width=3.4in, clip, keepaspectratio]{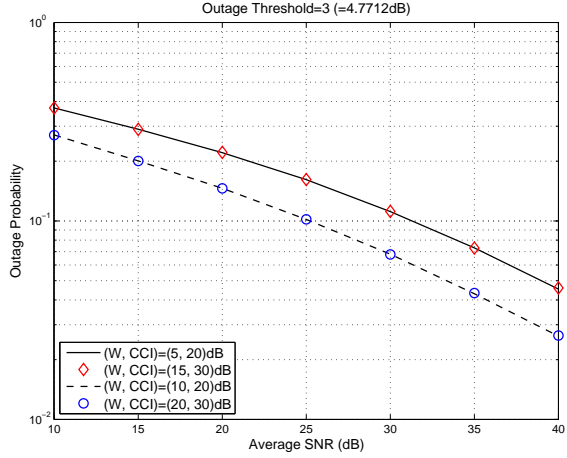}
\caption{Outage probability of the received SIR at the base station.}  \label{Fig.OutageUplink}
\end{figure}

Figure~\ref{Fig.OutageUplink} illustrates the outage probability at $\mathrm{BS}_1$ versus the average SIR in dB, where the values of both the average tolerable interference power $W$ and the CCI are varied. It is observed that the outage probability is determined only by the difference between $W$ and CCI, i.e., the value of $W-\mathrm{CCI}$. More specifically, it is clear to see that the cases with $(W,~\mathrm{CCI})=(5,~20)$ dB and with $(W,~\mathrm{CCI})=(15,~30)$ dB have the same outage probability whereas the cases with $(W,~\mathrm{CCI})=(10,~20)$ dB and with $(W,~\mathrm{CCI})=(20,~30)$ dB have the same outage probability. Moreover, the former's outage probability is larger than the latter's, because the former cases have smaller differences between $W$ and CCI than the latter cases (i.e., $-15$dB versus $-10$dB). In other words, larger difference between $W$ and CCI leads to lower outage probability. Intuitively speaking, this observation is not surprising because larger tolerable interference power allows higher transmit power at $\mathrm{SU}_1$, which in turn benefits mitigating the detrimental effect of higher CCI. Mathematically speaking, \eqref{Eq.CapacityD} implies that, for fixed CCI ($P$), larger tolerable interference power ($W$) leads to higher water-level of the optimal power-allocation at $\mathrm{SU}_1$ ($\lambda$). Moreover, \eqref{Eq.gamma2-R1} shows that $\gamma_2$ is determined by the ratio of $\lambda$ to CCI. In other words, $\gamma_2$ depends only upon the difference between $W$ and CCI in dB, and so does the received SIR  at $\mathrm{BS}_1$ ($\gamma_{\mathrm{bs_1}}$ given by \eqref{Eq.SIR@BS}), since $\gamma_{\mathrm{bs_1}}$ is a monotonically increasing function of $\gamma_2$.

\begin{figure}[t]
\centering
\includegraphics [width=3.4in, clip, keepaspectratio]{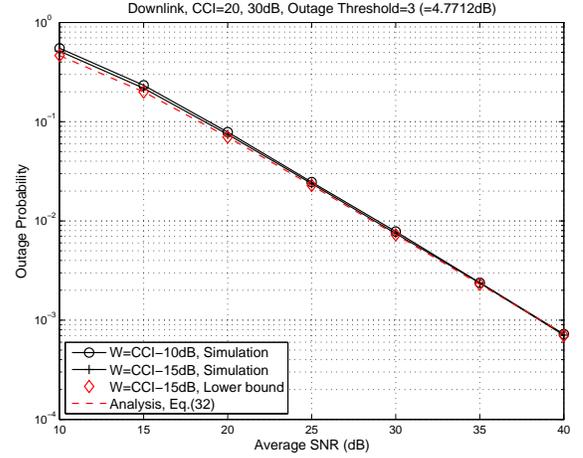}
\caption{Outage probability of the received SIR at the SU.}  \label{Fig.OutageDownlink}
\end{figure}

Now, we turn to the downlink from the base station $\mathrm{BS}_1$ to $\mathrm{SU}_1$. Figure~\ref{Fig.OutageDownlink} depicts the outage probability at $\mathrm{SU}_1$ versus the average SIR in dB, where the values of CCI are set to $20$ and $30$ dB while the values of average tolerable interference power $W$ are set to $\mathrm{CCI}-10$ dB and $\mathrm{CCI}-15$ dB. Like the observations in Fig.~\ref{Fig.OutageUplink}, the outage probability at $\mathrm{SU}_1$ is irrelevant to the actual values of $W$ and CCI but is determined only by their difference. Therefore, only three plots are shown in Fig.~\ref{Fig.OutageDownlink}. The upper plot corresponds to the case with $W=\mathrm{CCI}-10$ dB, the middle one refers to the case $W=\mathrm{CCI}-15$ dB, and the lower stands for the lower bound. It is seen that all simulation results are very tight with the lower bound, and the analytical results computed by \eqref{Eq.CDFSNR@SU-bound} coincide exactly with the simulation results. On the other hand, it is observed from Fig.~\ref{Fig.OutageDownlink} that, decreasing the values of $W$ will decrease the outage probability until the lower bound. This is because, decreasing the values of $W$ means lower transmit power at $\mathrm{SU}_1$. Then, as per \eqref{Eq.gamma3-5}, when the transmit power at $\mathrm{SU}_1$ approaches zero, $\gamma_5$ reduces to $\gamma_4$ and, hence, the received SIR at $\mathrm{SU}_1$ approaches its upper bound shown in \eqref{Eq.SNR@SU-bound}.

Finally, by comparing Fig.\ref{Fig.OutageUplink} with Fig.~\ref{Fig.OutageDownlink}, it is evident that the downlink performance of the considered relaying link is generally much better than the uplink. This is due to the strictly limited transmit power at $\mathrm{SU}_1$. This characteristics makes the proposed scheme particularly suitable for those wireless services with asymmetric traffics, such as e-mail checking, web browsing and video-on-demand, where the downlink traffic is much heavier than the uplink.

In summary, for the proposed two-way relaying in spectrum-sharing cellular systems, the uplink performance of secondary transmission (from a source SU to the BS via a PU serving as relaying node) is dominated by the difference between the average tolerable interference power at PUs and the CCI coming from concurrent primary transmission. Larger difference benefits improving uplink performance significantly in the whole SIR region of interest. On the other hand, the downlink (from the BS to the SU via a PU serving as relaying node) behaves like conventional one-way AF relaying and its performance is totally insensitive to the actual values of the average tolerable interference power at PUs and the CCI but is dominated by the average SIR.

\subsection{Effect of the Optimal Power Allocation on the Achievable Data Rate}
\begin{figure}[t]
\centering
\includegraphics [width=3.4in, clip, keepaspectratio]{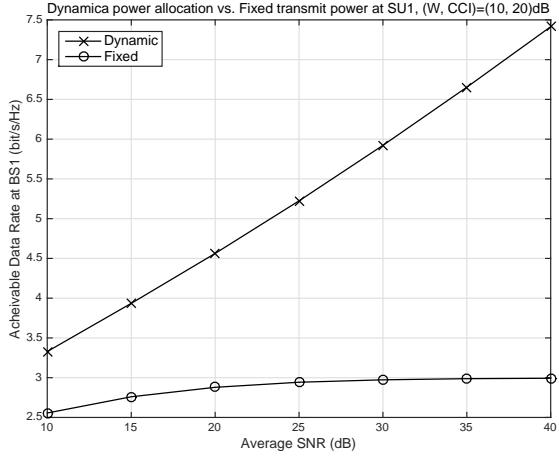}
\caption{Achievable data rates at $\mathrm{BS}_1$, with respect to different power allocation strategies at $\mathrm{SU}_1$.}
\label{Fig.SumRate}
\end{figure}

As shown in \eqref{Eq.CapacityA}--\eqref{Eq.CapacityD}, performing optimal power allocation at SUs is at the cost of additional CSI. In particular, the instantaneous CSI $u$ ($\mathrm{PU}_4 \to \mathrm{BS}_1$) and $v$ ($\mathrm{PU}_4 \to \mathrm{PU}_1$) shown in Fig.~\ref{Fig.SimulationSetting} have to be acquired prior to computing the optimal transmit power as per \eqref{Eq.CapacityC}. To illustrate of the benefits of the optimal power allocation, Fig.~\ref{Fig.SumRate} compares the achievable data rates at $\mathrm{BS}_1$ pertaining to the scenarios with the optimal power allocation and with a fixed transmit power at $\mathrm{SU}_1$ (i.e., the transmit power is only determined by \eqref{Eq.CapacityB}).

As shown by the upper curve with X-mark in Fig.~\ref{Fig.SumRate}, if the optimal power allocation is performed at $\mathrm{SU}_1$, the achievable data rate at $\mathrm{BS}_1$ increases in the whole SNR range of interest. On the contrary, if the transmit power at $\mathrm{SU}_1$ is fixed, the lower curve with circle marks illustrates that the achievable data rate at $\mathrm{BS}_1$ increases slightly in the low and medium SNR regime yet saturates at high SNR. The reason behind this observation is that the link performance is interference-limited by recalling the fact that the CCI is set to 20dB. More specifically, at $\mathrm{SNR} = 25$dB, the achievable data rate is about $5.2$ bit/s/Hz if the dynamic power allocation is performed at $\mathrm{SU}_1$ whereas it is only $2.9$ bit/s/Hz if the transmit power at $\mathrm{SU}_1$ is fixed. In other words, the optimal power allocation yields $1.8$ times higher data rate, compared with the strategy of fixed transmit power. Also, this data-rate gain becomes larger with higher SNR, as shown in Fig.~\ref{Fig.SumRate}. As a result, it is deducible that dynamic power allocation at SUs in spectrum-sharing cellular networks benefits effectively mitigating CCI and thus improving the achievable data rate of secondary transmission.

\section{Concluding Remarks}
\label{Section:Conclusion}
In this paper, spectrum-sharing technology is integrated into cellular networks by identifying potential secondary users. In order to guarantee the success of at most two-hop relaying transmission between a secondary user, which is out of femtocells and far from macrocell base station, and the macrocell base station, an idle primary user in the vicinity of the secondary user is chosen to serve as a relaying node. This new relaying paradigm differs completely from the conventional relaying strategies where only secondary users can assist the transmission of a source secondary user. By analyzing the outage probability of the proposed dual-hop two-way relaying scheme, it is revealed that the downlink performance from the base station to the secondary user outperforms the uplink performance from the secondary user to the base station, both via the relaying node. This asymmetric downlink/uplink performance makes the proposed relaying scheme particularly suitable for the wireless services where the downlink traffic is much heavily than the uplink, such as e-mail checking, web browsing, social networking and data streaming. Actually, these services are most attractive to secondary users.

\begin{appendix}[Proof of Proposition~\ref{Proposition1@SU}]
In light of the definitions of $\gamma_1$, $\gamma_3$ and $\gamma_4$ shown in \eqref{Eq.gamma1-2}-\eqref{Eq.gamma3-5}, the distribution functions of $\gamma_3$ and $\gamma_4$ are similar to those of $\gamma_1$ given by \eqref{Eq.PDFgamma1}-\eqref{Eq.CDFgamma1}. More specifically,
\begin{eqnarray}
f_{\gamma_{_3}}(x) = \eta_2\bar{\gamma}\left(x+\eta_2\bar{\gamma}\right)^{-2}, \label{Eq.PDFgamma3} \\
F_{\gamma_{_3}}(x) = 1-\eta_2\bar{\gamma}\left(x+\eta_2\bar{\gamma}\right)^{-1}; \label{Eq.CDFgamma3} \\
f_{\gamma_{_4}}(x) = \eta_3\bar{\gamma}\left(x+\eta_3\bar{\gamma}\right)^{-2}, \label{Eq.PDFgamma4} \\
F_{\gamma_{_4}}(x) = 1-\eta_3\bar{\gamma}\left(x+\eta_3\bar{\gamma}\right)^{-1}. \label{Eq.CDFgamma4}
\end{eqnarray}
With the distribution functions of $\gamma_3$ and $\gamma_4$ developed, the CDF of $\gamma^{\prime}_{\mathrm{su}_1}$ shown in \eqref{Eq.SNR@SU-bound} can be expressed as
\begin{eqnarray}
\lefteqn{\mathrm{Pr}\left\{\gamma^{\prime}_{\mathrm{su}_1} < \gamma\right\}}        \nonumber \\
&  =  & \mathcal{E}\left\{\mathrm{Pr}\left(\frac{\gamma_3\gamma_4}{\gamma_3+\gamma_4} < \gamma \vert \gamma_4\right)\right\}  \nonumber \\
&  =  & \mathcal{E}\left\{\mathrm{Pr}\left(\gamma_3(\gamma_4-\gamma) < \gamma\gamma_4 \vert \gamma_4\right)\right\}  \nonumber \\
&  =  & \int_{\gamma}^\infty{\mathrm{Pr}\left(\gamma_3 < \frac{\gamma\gamma_4}{\gamma_4-\gamma}\right)f_{\gamma_{_4}}(\gamma_4)}\,\mathrm{d}\gamma_4  \nonumber \\
&     & {} + \int_0^{\gamma}{\mathrm{Pr}\left(\gamma_3 > \frac{\gamma\gamma_4}{\gamma_4-\gamma}\right)f_{\gamma_{_4}}(\gamma_4)}\,\mathrm{d}\gamma_4  \nonumber \\
&  =  & 1-F_{\gamma_4}(\gamma) -
\underbrace{\int_{\gamma}^\infty{\mathrm{Pr}\left(\gamma_3 > \frac{\gamma\gamma_4}{\gamma_4-\gamma}\right)f_{\gamma_{_4}}(\gamma_4)}\,\mathrm{d}\gamma_4}_{I_1}  \nonumber \\
&     & {} + \underbrace{\int_0^{\gamma}{\mathrm{Pr}\left(\gamma_3 > \frac{\gamma\gamma_4}{\gamma_4-\gamma}\right)f_{\gamma_{_4}}(\gamma_4)}\,\mathrm{d}\gamma_4}_{I_2}.
\label{Appendix-B-1}
\end{eqnarray}
Then, we derive the integral terms $I_1$ and $I_2$ in closed-form.

By virtue of  the CDF of $\gamma_3$ shown in \eqref{Eq.CDFgamma3} and the PDF of $\gamma_4$ given by \eqref{Eq.PDFgamma4}, the integral term $I_1$ shown in \eqref{Appendix-B-1} can be rewritten as
\begin{eqnarray}
I_1
&  =  & \int_{\gamma}^\infty{\frac{\eta_2\bar{\gamma}}{\left(\frac{\gamma\gamma_4}{\gamma_4-\gamma}+\eta_2\bar{\gamma}\right)} \times
\frac{\eta_3\bar{\gamma}}{\left(\gamma_4+\eta_3\bar{\gamma}\right)^2}}\,\mathrm{d}\gamma_4 		 \nonumber\\
&  =  & \frac{\eta_2\eta_3\bar{\gamma}^2}{\gamma+\eta_2\bar{\gamma}}\int_{0}^\infty{\frac{x}{\left(x+\frac{\gamma^2}{\gamma+\eta_2\gamma}\right)\left(x+\gamma+\eta_3\bar{\gamma}\right)^2}
}\,\mathrm{d}x 		 \nonumber\\
&  =  &\frac{\eta_2\eta_3\bar{\gamma}^2\gamma^2}{2(\gamma+\eta_2\bar{\gamma})^2(\gamma+\eta_3\bar{\gamma})^2}  \nonumber \\
&      &{}\times{_2F_1}\left(2, 2; 3; 1-\frac{\gamma^2}{(\gamma+\eta_2\bar{\gamma})(\gamma+\eta_3\bar{\gamma})}\right)
\label{Appendix-B-2}
\end{eqnarray}
where \cite[vol.1, Eq.(2.2.6.24)]{Prudnikov86} was exploited to attain \eqref{Appendix-B-2}, with ${_2F_1}( \cdot\,, \cdot\,; \cdot\,; x)$ being the Gaussian hypergeometric function \cite[Eq.(15.2.1)]{NIST10}.

As far as the integral term $I_2$ in \eqref{Appendix-B-1} is concerned, it is clear that $\mathrm{Pr}\left\{\gamma_3 > \frac{\gamma\gamma_4}{\gamma_4-2\gamma}\right\} = 1$ since $\gamma_4 < 2\gamma$. Therefore, $I_2$ can be easily computed by
\begin{equation} \label{Appendix-B-3}
I_2
= \int_0^{\gamma}{f_{\gamma_{_2}}(\gamma_4)}\,\mathrm{d}\gamma_4
= F_{\gamma_4}(\gamma).
\end{equation}
Then, substituting \eqref{Appendix-B-2} and \eqref{Appendix-B-3} into \eqref{Appendix-B-1} yields the desired CDF expression shown in \eqref{Eq.CDFSNR@SU-bound}. Finally, by recalling the first-order derivative of the Gaussian hypergeometric function, i.e., $\frac{\mathrm{d}}{\mathrm{d}x}{_2F_1(a, b; c; x)} = \frac{ab}{c}{_2F_1(a+1, b+1; c+1; x)}$ \cite[Eq.(15.5.1)]{NIST10} and taking the derivative of \eqref{Eq.CDFSNR@SU-bound} with respect to $x$ as well as performing some algebraic manipulations, it is not hard to attain the PDF given by \eqref{Eq.PDFSNR@SU-bound}.
\end{appendix}

\begin{IEEEbiography}[{\includegraphics[width=1in, height=1.25in,  clip, keepaspectratio]{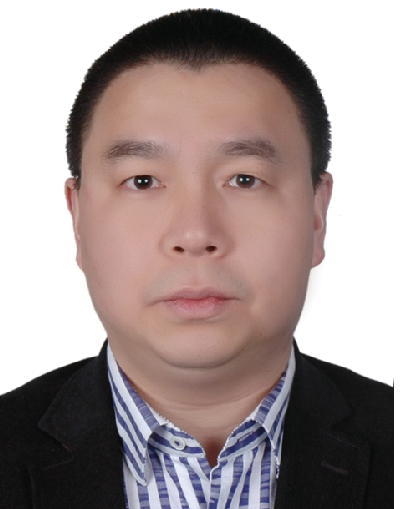}}]{Minghua Xia} (M'12) obtained his Ph.D. degree in Telecommunications and Information Systems from Sun Yat-sen University, Guangzhou, China, in 2007. Since 2015, he has been working as a Professor at the same university.

From 2007 to 2009, he was with the Electronics and Telecommunications Research Institute (ETRI) of South Korea, Beijing R\&D Center, Beijing, China, where he worked as a member and then as a senior member of engineering staff and participated in the projects on the physical layer design of 3GPP LTE mobile communications. From 2010 to 2014, he was in sequence with The University of Hong Kong, Hong Kong, China; King Abdullah University of Science and Technology, Jeddah, Saudi Arabia; and the Institut National de la Recherche Scientifique (INRS), University of Quebec, Montreal, Canada, as a Postdoctoral Fellow. His research interests are in the general area of 5G wireless communications, and in particular the design and performance analysis of multi-antenna systems, cooperative relaying systems and cognitive relaying networks, and recently focus on the design and analysis of wireless power transfer and/or energy harvesting systems, as well as massive MIMO and small cells. He holds two patents granted in China.

Dr. Xia received the Professional Award at IEEE TENCON'15, Macau, 2015. He was also awarded as an Exemplary Reviewer by {\scshape IEEE Transactions on Communications}, {\scshape IEEE Communications Letters}, and {\scshape IEEE Wireless Communications Letters}, respectively, in 2014.
\end{IEEEbiography}

\vfill

\begin{IEEEbiography}
[{\includegraphics[width=1in, height=1.25in,  clip, keepaspectratio]{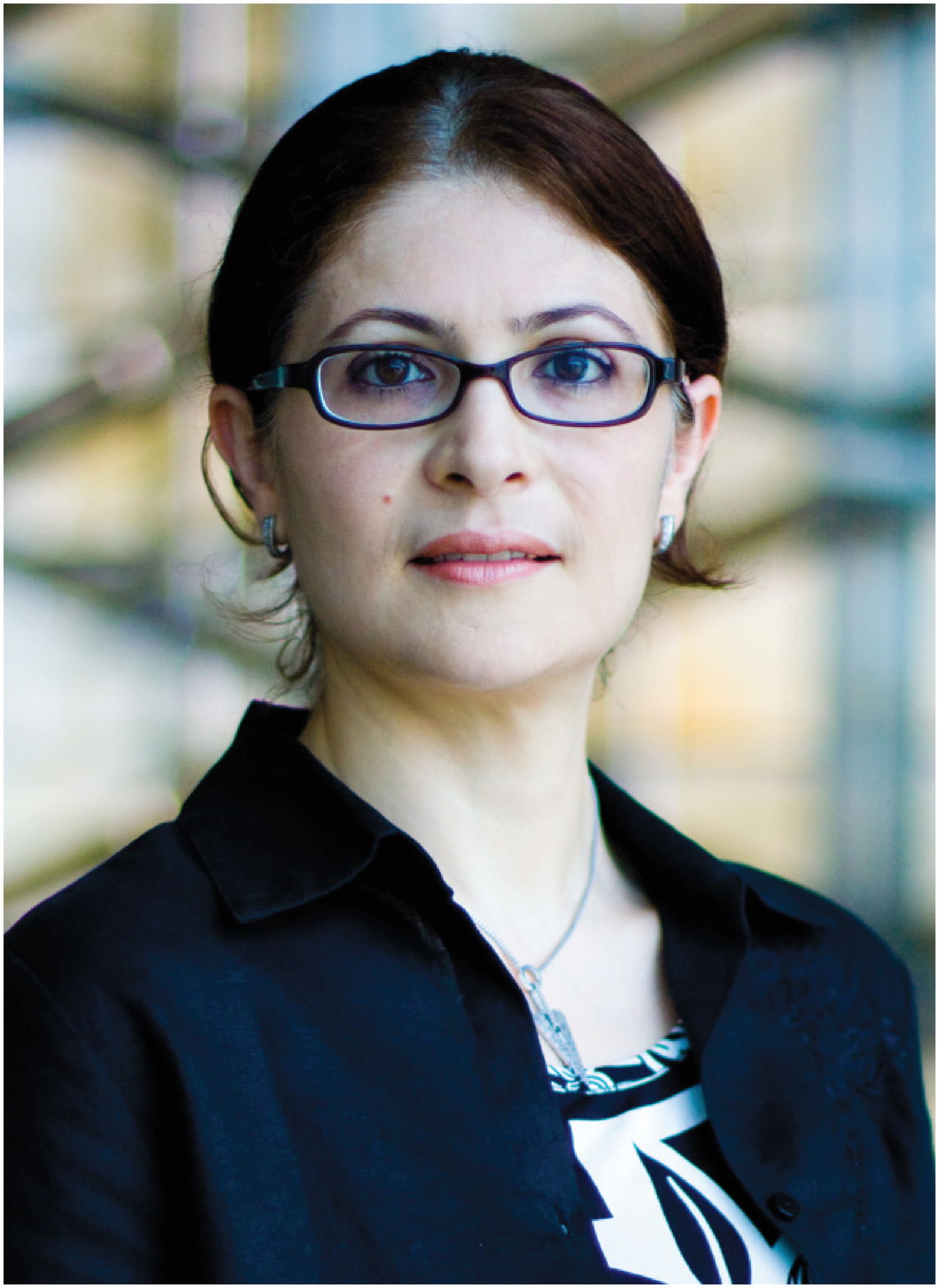}}]{Sonia A\"{\i}ssa} (S'93-M'00-SM'03) received her Ph.D. degree in Electrical and Computer Engineering from McGill University, Montreal, QC, Canada, in 1998. Since then, she has been with the Institut National de la Recherche Scientifique-{\it Energy, Materials and Telecommunications} Center (INRS-EMT), University of Quebec, Montreal, QC, Canada, where she is a Full Professor.

From 1996 to 1997, she was a Researcher with the Department of Electronics and Communications of Kyoto University,
and with the Wireless Systems Laboratories of NTT, Japan. From 1998 to 2000, she was a Research Associate at INRS-EMT. In 2000-2002, while she was an Assistant Professor, she was a Principal Investigator in the major program of personal and mobile communications of the Canadian Institute for Telecommunications Research, leading research in radio resource management for wireless networks. From 2004 to 2007, she was an Adjunct Professor with Concordia University, Montreal. She was Visiting Invited Professor at Kyoto University, Japan, in 2006, and Universiti Sains Malaysia, in 2015.
Her research interests include the modeling, design and performance analysis of wireless communication systems and networks.

Dr. A\"{\i}ssa is the Founding Chair of the IEEE Women in Engineering Affinity Group in Montreal, 2004-2007; acted as TPC Symposium Chair or Cochair at IEEE ICC '06 '09 '11 '12; Program Cochair at IEEE WCNC 2007; TPC Cochair of IEEE VTC-spring 2013; and TPC Symposia Chair of IEEE Globecom 2014. Her main editorial activities include: Editor, {\scshape IEEE Transactions on Wireless Communications}, 2004-2012; Associate Editor and Technical Editor, {\scshape IEEE Communications Magazine}, 2004-2015; Technical Editor, {\scshape IEEE Wireless Communications Magazine}, 2006-2010; and Associate Editor, {\it Wiley Security and Communication Networks Journal}, 2007-2012. She currently serves as Area Editor for the {\scshape IEEE Transactions on Wireless Communications}. Awards to her credit include the NSERC University Faculty Award in 1999; the Quebec Government FRQNT Strategic Faculty Fellowship in 2001-2006; the INRS-EMT Performance Award multiple times since 2004, for outstanding achievements in research, teaching and service; and the Technical Community Service Award from the FQRNT Centre for Advanced Systems and Technologies in Communications, 2007. She is co-recipient of five IEEE Best Paper Awards and of the 2012 IEICE Best Paper Award; and recipient of NSERC Discovery Accelerator Supplement Award. She is a Distinguished Lecturer of the IEEE Communications Society (ComSoc) and an Elected Member of the ComSoc Board of Governors. Professor A\"{\i}ssa is a Fellow of the Canadian Academy of Engineering.
\end{IEEEbiography}

\vfill


\begin{thebibliography}{99}
\bibitem{Srinivasa07Mag05}
S.~Srinivasa and S.~A.~Jafar,  ``The throughput potential of cognitive radio: a theoretical perspective,'' \emph{IEEE Commun. Mag.}, vol.~45, no.~5, pp.~73--79, May 2007.

\bibitem{MusavianIET08}
L. Musavian and S.~A\"{i}ssa, ``Outage-constrained capacity of spectrum-Sharing channels in fading environments,'' {\em IET Commun., Special Issue on Cognitive Spectrum Access}, vol. 2, no. 6, pp. 724--732, July 2008.

\bibitem{LeeTWC11Feb}
J.~Lee, H.~Wang, J.~G.~Andrews, and D.~Hong, ``Outage probability of cognitive relay networks with interference constraints,'' \emph{IEEE Trans. Wireless Commun.}, vol.~10, no.~2, pp.~390--395, Feb.~2011.

\bibitem{AsghariICC10}
V. Asghari and S.~A\"{i}ssa, ``Cooperative relay communication performance under spectrum-sharing resource requirements,'' in {\em Proc. IEEE ICC'10}, pp. 1--6, Cape Town, South Africa, May 2010.

\bibitem{XiaTCOM12June}
M.~Xia and S.~A\"{i}ssa, ``Cooperative AF relaying in spectrum-sharing systems: performance analysis under average interference power constraints and Nakagami-$m$ fading,'' \emph{IEEE Trans. Commun.}, vol.~60, no.~6, pp.~1523--1533, June~2012.

\bibitem{XiaWCL14}
M. Xia and S. A\"{i}ssa, ``Impact of co-channel interference on the performance of multi-hop relaying over Nakagami-$m$ fading channels,'' \emph{IEEE Wireless Commun. Lett.}, vol.~3, no.~2, pp.~133--136, Apr. 2014.

\bibitem{AndrewsJSAC1403}
J. G. Andrews, H. Claussen, M. Dohler, S. Rangan, and M. C. Reed, ``Femtocells: past, present, and future,'' \emph{IEEE J. Select. Area Commun.}, vol.~30, no.~3, pp.~497--508, Apr.~2012.

\bibitem{ChandrasekharCM0809}
V. Chandrasekhar, J. G. Andrews, and A. Gatherer, ``Femtocell networks: a survey,'' \emph{IEEE Commun. Mag.}, vol.~46, no.~9, pp.~59--67, Sep. 2008.

\bibitem{Xia13Mag12}
M. Xia and S. A\"{i}ssa, ``Underlay cooperative AF relaying in cellular networks: performance and challenges,'' \emph{IEEE Commun. Mag.}, vol.~51, no.~12, pp.~170--176, Dec. 2013.

\bibitem{NTIA11}
Final report of the Incentives Subcommittee of Spectrum Management Advisory Committee of NTIA, Jan. 11, 2011. Available: \url{http://www.ntia.doc.gov/report/2011/final-report-incentives-subcommittee}.

\bibitem{PCAST12}
The report to the President by the President's Council of Advisors on Science and Technology (PCAST), ``Realizing the full potential of government-held spectrum to spur economic growth,'' July 2012, Available: \url{http://www.whitehouse.gov/sites/default/files/microsites/ostp/pcast_spectrum_report_final_july_20_2012.pdf}.

\bibitem{Nasab13TCOM08}
E. Soleimani-Nasab, M. Matthaiou, M. Ardebilipour, and G. K. Karagiannidis, ``Two-way AF relaying in the presence of co-channel interference'', \emph{IEEE Trans. Commun.}, vol.~61, no.~8, pp.~3156--3169, Aug.~2013.

\bibitem{RankovJSAC0702}
B. Rankov and A. Wittneben, ``Spectral efficient protocols for half-duplex fading relaying channels,'' \emph{IEEE J. Sel. Areas Commun.}, vol. 25, no. 2, pp. 373--389, Feb. 2007.

\bibitem{KnoppICC95}
R.~Knopp and P.~A.~Humblet, ``Information capacity and power control in single-cell multiuser communications,'' in \emph{Proc. IEEE ICC'95}, vol.~1, pp.~331--335, Seattle, USA, 1995.

\bibitem{XuTVT14}
K.~Xu, D.~Zhang, Y.~Xu, and W.~Ma, ``On the equivalence of two optimal power allocation schemes for A-TWRC,'' \emph{IEEE Trans. Veh. Tech.}, vol.~63, no.~4, pp.~1970--1976, May 2014.

\bibitem{Tse05}
D.~Tse and P. ~Wiswanath, \emph{Fundamentals of Wireless Communications}, Cambridge University Press, 2005.

\bibitem{GoldsmithTIT97Nov}
A.~J.~Goldsmith and P.~P.~Varajya, ``Capacity of fading channels with channel side information,'' \emph{IEEE Trans. Inf. Theory}, vol.~43, no.~6, pp.~1986--1992, Nov.~1997.

\bibitem{MusavianTWC09Jan}
L.~Musavian and S.~A\"{i}ssa, ``Capacity and power allocation for spectrum-sharing communications in fading channels,'' \emph{IEEE Trans. Wireless Commun.}, vol.~8, no.~1, pp.~148--156, Jan.~2009.

\bibitem{Zhang09TWC04}
R.~Zhang, ``On peak versus average interference power constraints for protecting primary users in cognitive radio networks,'' \emph{IEEE Trans. Wireless Commun.}, vol.~8, no.~4, Apr.~2009, pp.~2112--2120.

\bibitem{Balakrishnan03}
N. Balakrishnan and V. B. Nevzorov, \emph{A Primer on Statistical Distributions}, John Wiley \& Sons Inc., 2003.

\bibitem{NIST10}
F.~W.~J.~Olver, \emph{NIST Handbook of Mathematical Functions}, Cambridge University Press, 2010.

\bibitem{Prudnikov86}
A.~P.~Prudnikov, Y.~A.~Brychkov, and O.~I.~Marichev, \emph{Integrals and Series}, Gordon and Breach Science Publishers, 1986.

\bibitem{AnghelTWC04May}
P.~A.~Anghel and M.~Kaveh, ``Exact symbol error probability of a cooperative network in a Rayleigh-fading environment,'' \emph{IEEE Trans. Wireless Commun.}, vol.~3, no.~5, pp.~1416--1421, May~2004.

\bibitem{Andrews11TCOM11}
G.~Andrews, F.~Baccelli, and R.~K.~Ganti, ``A tractable approach to coverage and rate in cellular networks,'' \emph{IEEE Trans. Commun.}, vol.~59, no.~11, pp.~3122--3134, Nov. 2011.
\end{thebibliography}
\end{document}